\documentclass{article}
\usepackage{amsmath,amssymb,bbm,color,theorem}

\newcommand{\email}[1]{{\textit{Email:} \texttt{#1}}}
\newcommand{\subindex}[2]{\index{#1!#2}}
\newcommand{\subsubindex}[3]{\index{#1!#2!#3}}
\newcommand{\tmem}[1]{{\em #1\/}}
\newcommand{\tmop}[1]{\ensuremath{\operatorname{#1}}}
\newcommand{\tmtextbf}[1]{{\bfseries{#1}}}
\newcommand{\tmtextit}[1]{{\itshape{#1}}}
\newcommand{\tmtextrm}[1]{{\rmfamily{#1}}}
\newcommand{\tmtextup}[1]{{\upshape{#1}}}
\newcommand{\um}{-}

\newenvironment{proof*}[1]{\noindent\textbf{#1\ }}{\hspace*{\fill}$\Box$\medskip}

\definecolor{grey}{rgb}{0.75,0.75,0.75}
\definecolor{orange}{rgb}{1.0,0.5,0.5}
\definecolor{brown}{rgb}{0.5,0.25,0.0}
\definecolor{pink}{rgb}{1.0,0.5,0.5}
{\theorembodyfont{\rmfamily}\newtheorem{remark}{Remark}}
\newtheorem{theorem}{Theorem}

\begin{document}

\title{Bounding the radii of balls meeting every connected component
of semi-algebraic sets}
\author{Saugata Basu
\thanks{\email{sbasu@math.purdue.edu}}\\
Department of Mathematics\\
Purdue University\\
West Lafayette, IN 47907\\
USA\\
\and
Marie-Fran\c{c}oise Roy
\thanks{\email{marie-francoise.roy@univ-rennes1.fr}}\\
IRMAR (URA CNRS 305)\\
Universite de Rennes 1\\
Campus de Beaulieu\\
35042 Rennes, cedex\\
France}
\maketitle

\begin{abstract}
  We prove explicit bounds on the radius of a ball centered at the origin
  which is guaranteed to contain all bounded connected components of a
  semi-algebraic set $S \subset \mathbbm{R}^k$ defined by a quantifier-free
  formula involving $s$ polynomials in \ $\mathbbm{Z}[X_1, \ldots, X_k]$
  having degrees at most $d$, and whose coefficients have bitsizes at most
  $\tau$. Our bound is an explicit function of $s, d, k$ and $\tau$, \ and
  does not contain any undetermined constants. We also prove a similar bound
  on the radius of a ball guaranteed to intersect every connected component of
  $S$ (including the unbounded components). While asymptotic bounds of the
  form \ $2^{\tau d^{O (k)}}$ on these quantities were known before, some
  applications require bounds which are explicit and which hold for all values
  of $s, d, k$ and $\tau$. The bounds proved in this paper are of this nature.
  
\end{abstract}

\section{Introduction }

Let $S \subset \mathbbm{R}^k$ be a semi-algebraic subset of $\mathbbm{R}^k$
defined by a quantifier-free formula whose atoms are of the form $P \left\{ >,
<, =\}0, P \in \mathcal{P} \right.$, where \ $\mathcal{P} \subset
\mathbbm{Z}[X_1, \ldots, X_k]$ is a set of polynomials, with $\#\mathcal{P}=
s$, $\deg (P) \leq d$ for $P \in \mathcal{P}$, and the bitsizes of the
coefficients of $P \in \mathcal{P}$ are bounded by $\tau$. In this paper we
consider the problem of obtaining an upper bound on the radius of a ball
guaranteed to {\tmem{contain}} all {\tmem{bounded}} semi-algebraically
connected components of $S$, as well as on the radius of a ball guaranteed to
{\tmem{meet}} every semi-algebraically connected component of $S$. Such bounds
have many applications in different areas of mathematics as well as computer
science. For instance, bounds of these types play a critical role in recent
work on proving uniform bounds in the infinitesimal version of Hilbert's
sixteenth problem {\cite{BNY08,BYak08}}, as well as in proving certain lower
bounds in computer science {\cite{HKM09}}.

We obtain explicit upper bounds (in terms of $s, d, k$ and $\tau$) on the
radii of such balls in each of the two cases mentioned above. Our bounds are
slightly better in the special case when the semi-algebraic set $S$ is a real
algebraic variety defined by one polynomial equation (in this case $s = 1$).
Indeed, the bound in the \ general case is proved by reducing the problem to
this special case. Hence, we first prove the results for algebraic sets in
Section \ref{sec:algebraic}, and prove the bounds for general semi-algebraic
sets in Section \ref{sec:semi-algebraic}.

\subsection{History }

Asymptotic bounds on the radius of a ball guaranteed to meet all connected
components of a semi-algebraic subset of $\mathbbm{R}^k$ defined by a
quantifier-free formula involving polynomials in $\mathbbm{Z}[X_1, \ldots,
X_k]$ in terms of the number $s$, the maximum degree $d$, and the maximum
bitsize $\tau$ of the coefficents of the defining polynomials, were known
before. The best of these bounds were of the form $2^{\tau d^{O (k)}}$
{\cite{BPRbook2posted,GV,R92}}, with undetermined constants, and it seems that
there is little hope to improve them in a significant way. While such bounds
are already useful in many contexts, certain applications might require more
precise and completely explicit estimates valid for all values of $s, d, k,$
and $\tau$. This is what we do in this paper.

\section{Main Results}

\subsection{Some notation}

We first fix some notation.

Let $\text{\tmtextrm{\tmtextup{R}}}$ be a real closed field. If $\mathcal{P}$
is a finite subset of $\text{\tmtextrm{\tmtextup{R}}} [X_1, \ldots, X_k]$, we
write the \tmtextbf{set of zeros} of $\mathcal{P}$ in
$\text{\tmtextrm{\tmtextup{R}}}^k$ as
\[ \tmop{Zer} ( \mathcal{P}, \text{\tmtextrm{\tmtextup{R}}}^k) =\{x \in
   \text{\tmtextrm{\tmtextup{R}}}^k \mid \bigwedge_{P \in \mathcal{P}} P (x) =
   0\}. \]
A \tmtextbf{sign condition} on $\mathcal{\mathcal{P}}$ is an element of $\{0,
1, \um 1\}^{\mathcal{\mathcal{P}}}$, i.e. a mapping from $\mathcal{P}$ to
$\{0, 1, - 1\}$.

We say that $\mathcal{\mathcal{P}}$ \tmtextbf{realizes} the sign condition
$\sigma$ at $x \in \text{\tmtextrm{\tmtextup{R}}}^k$ if~$\bigwedge_{P \in
\mathcal{\mathcal{P}}} \tmop{sign} (P (x)) = \sigma (P) .$

The \tmtextbf{realization of the sign condition $\sigma$}
\[ \tmop{Reali} (\sigma) =\{x \in \text{\tmtextrm{\tmtextup{R}}}^k
   \hspace{0.75em} \mid \hspace{0.75em} \bigwedge_{P \in
   \mathcal{\mathcal{P}}} \tmop{sign} (P (x)) = \sigma (P)\}. \]
The sign condition $\sigma$ is \tmtextbf{realizable} if $\tmop{Reali}
(\sigma)$ is non-empty.

Given an integer $n$, we denote by $\tmop{bit} (n)$ the number of bits of its
absolute value in the binary representation.

The main results of the paper can now be stated as follows. Our results in
the algebraic case are slightly better than in the semi-algebraic case and we
state them separately.

\subsection{Algebraic Case}

\begin{theorem}
  \label{12:the:algebraicsmallball}(Ball containing all bounded components)
  Let $Q \in \mathbbm{Z}[X_1, \ldots, X_k]$ be a polynomial with $\deg (Q) =
  d$, and suppose that the coefficients of $Q$ in $\mathbb{Z}$ have bitsizes
  at most $\tau$. Then, every bounded semi-algebraically connected component
  of $\tmop{Zer} (Q, \mathbbm{R}^k)$ is contained inside a ball centered at
  the origin of radius
  \[ k^{1 / 2} (N + 1) 2^{N (kd + 2) (\tau + \tmop{bit} (N) + \tmop{bit} (d +
     1))} \]
  where
  \[ N = (d + 1) d^{k - 1} . \]
  In particular, all isolated points of $\tmop{Zer} (Q, \mathbbm{R}^k)$ are
  contained inside the same ball.
\end{theorem}

\begin{theorem}
  \label{12:the:algebraicball}(Ball meeting all components) Let $Q \in
  \mathbbm{Z}[X_1, \ldots, X_k]$ be a polynomial of degree with $\deg (Q) = d$
  and suppose that \ the coefficients of $Q$ in $\mathbb{Z}$ have bitsizes at
  most $\tau$. Then there exists a ball centered at the origin of radius
  bounded by
  \[ \left( (2 D N (2 N - 1) + 1) 2^{\left. (2 N - 1) \tau' + N^2 \tmop{bit}
     (N + 1 \right)} \right)^{1 / 2} \]
  intersecting every semi-algebraically connected component of $\tmop{Zer} (Q,
  \mathbbm{R}^k)$, where
  \begin{eqnarray*}
    D & = & k d' - 2 (k - 1),\\
    N & = & d' (d' - 1)^{k - 1},\\
    \tau' & = & 2 ND \tau + N (\rho + \rho'),
  \end{eqnarray*}
  with
  \begin{eqnarray*}
    d' & = & \sup (2 (d + 1), 6),\\
    \rho & = & D (k \tmop{bit} (d + 1) + \tmop{bit} (d') + 1 + 4 \tmop{bit} (2
    D + 1) + \tmop{bit} (N)) - 2 \tmop{bit} (2 D + 1),\\
    \rho' & = & (2 k - 2) \tmop{bit} (N) + k \tmop{bit} (k) + 2 \tmop{bit} (2
    DN + 1) + 1.
  \end{eqnarray*}
\end{theorem}

\subsection{Semi-algebraic case}

\begin{theorem}
  \label{13:the:diam}Given a set $\mathcal{P}=\{P_1, \ldots, P_s \} \subset
  \mathbbm{Z}[X_1, \ldots, X_k]$ with $\deg (P) \leq d, P \in \mathcal{P}$,
  such that the bitsizes of the coefficients of $P \in \mathcal{P}$ are
  bounded \ by $\tau$, there exists a ball centered at the origin of radius
  bounded by
  \[ k^{1 / 2} (N + 1) 2^{N (2 kd + 2) (2 \tau + \tmop{bit} (N) + (k + 1)
     \tmop{bit} (d + 1) + \tmop{bit} (s))}, \]
  where
  \[ N = {\color{black} (2 d + 1) (2 d)^{k - 1},} \]
  \[  \]
  which contains every bounded semi-algebraically connected component of the
  realization of every realizable sign condition on $\mathcal{P}$. 
\end{theorem}

\begin{theorem}
  \label{13:the:saball}Given a set $\mathcal{P}=\{P_1, \ldots, P_s \} \subset
  \mathbbm{Z}[X_1, \ldots, X_k]$ with $\deg (P) \leq d, P \in \mathcal{P}$,
  such that the bitsizes of the coefficients of $P \in \mathcal{P}$ are
  bounded \ by $\tau$, there exists a ball centered at the origin of radius
  \[ \left( (2 D N (2 N - 1) + 1) 2^{\left. (2 N - 1) \tau'' + N^2 \tmop{bit}
     (N + 1 \right)} \right)^{1 / 2} \]
  intersecting every semi-algebraically connected component of $\tmop{Zer} (Q,
  \mathbbm{R}^k)$, where
  \begin{eqnarray*}
    D & = & k d' - 2 (k - 1),\\
    N & = & d' (d' - 1)^{k - 1},\\
    \tau'' & = & 2 ND (2 \tau + k \tmop{bit} (d + 1) + \tmop{bit} (s)) + N
    (\rho + \rho'),
  \end{eqnarray*}
  with
  \begin{eqnarray*}
    d' & = & \sup (2 (d + 1), 6),\\
    \rho & = & D (k \tmop{bit} (d + 1) + \tmop{bit} (d') + 1 + 4 \tmop{bit} (2
    D + 1) + \tmop{bit} (N)) - 2 \tmop{bit} (2 D + 1),\\
    \rho' & = & (2 k - 2) \tmop{bit} (N) + k \tmop{bit} (k) + 2 \tmop{bit} (2
    DN + 1) + 1,
  \end{eqnarray*}
  which meets every semi-algebraically connected component of the realization
  of every realizable sign condition on $\mathcal{P}$. 
\end{theorem}

\begin{remark}
  Note that all the bounds above are of the form \ $2^{\tau d^{O (k)}}$,
  similarily to the results obtained in \ {\cite{BPRbook2posted,GV,R92}}. The
  only point which needs some explanation is the fact that $s$ plays a role in
  our estimates for the semi-algebraic case, while it does not appear in the
  formula \ $2^{\tau d^{O (k)}}$. This is because the total number of
  polynomials of degree $d$ in $k$ variables with bitsize bounded by $\tau$ is
  bounded by $(2^{\tau})^{\binom{d + k}{k}} = 2^{\tau d^{O (k)}}$.
\end{remark}

\section{Preliminaries}

In order to prove the bounds on the radii of various balls we need a careful
analysis of the bit sizes of the entries of certain matrices corresponding to
multiplication by certain variables in a zero-dimensional ideal of a very
special type. This analysis appearing in {\cite{BPRbook2posted}} is similar in
spirit to the techniques in {\cite{JP}}. We reproduce here the results
(without proofs which appear in {\cite{BPRbook2posted}}) for the benefit of
the readers.

Let $\text{\tmtextrm{\tmtextup{D}}}$ be an ordered domain. We first define a
special type of Groebner basis with coefficients in
$\text{\tmtextrm{\tmtextup{D}}}$. We say that $\mathcal{G} (Y, Z)$ is a
\tmtextbf{parametrized special Groebner basis}{\subindex{Gr\~A¶bner
basis}{special parametrized -}} if it is of the form

\[ \mathcal{G} (Y, Z) =\{ZX_1^{d_1} + Q_1 (Y, X), \ldots, ZX_k^{d_k} + Q_k (Y,
   X)\} \]

with $Q_i \in \text{\tmtextrm{\tmtextup{D}}} [Y] [X_1, \ldots, X_k], \deg
(Q_i) < d_i, \deg_{X_j} (Q_i) < d_j, i \not{=} j$, where $\deg$ is the total
degree with respect to the variables~$X_1, \ldots, X_k$, $d_1 \ge \ldots \ge
d_k \geq 1$, and $Z$ is either a new variable or one of the variables $Y_1,
\ldots, Y_{\ell}$. \ Define $\overline{\tmop{Mon}} ( \mathcal{G}) (Z)$ as the
set of elements $Z^{| \alpha |} X^{\alpha} = Z^{| \alpha |} X_1^{\alpha_1}
\cdots X_k^{\alpha_k}$ with~$\alpha_i < d_i$ and $\overline{\tmop{Bor}} (
\mathcal{G}) (Z)$ as the set of elements $Z^{| \alpha |} X^{\alpha}$ such that
$\alpha_i = d_i$ for some $i \in \{1, \ldots, k\}$ and $\alpha_i \le d_i$ for
any $i \in \{1, \ldots, k\}$.

The following algorithm is described in {\cite{BPRbook2posted}}. Here we just
recall the imput, output and the estimates on the bitsizes of the output.

{\noindent}\tmtextbf{Algorithm 1. } \tmtextbf{\tmtextup{[Parametrized Special
Matrices of Multiplication]}}\label{12:alg:pmult}{\subsubindex{Matrix of
multiplication}{Algorithm}{Parametrized Special}}
\begin{itemize}
  \item \tmtextbf{Structure:} a ring $\text{\tmtextrm{\tmtextup{D}}}$
  contained in a field $\text{\tmtextrm{\tmtextup{K}}}$.
  
  \item \tmtextbf{Input:} a parametrized special Grobner basis
  
  \[ \mathcal{G} =\{ZX_1^{d_1} + Q_1 (Y, X), \ldots, ZX_k^{d_k} + Q_k (Y, X)\}
     \subset \text{\tmtextrm{\tmtextup{D}}} [Y, Z] [X_1, \ldots, X_k] \]
  
  with $Y = (Y_1, \ldots, Y_{\ell})$.
  
  \item \tmtextbf{Output:} parametrized matrices of multiplication by the
  variables in the basis $\overline{\tmop{Mon}} ( \mathcal{G}) (Z)$: i.e. for
  every variable $X_i$ the matrix $M'_i (Y, Z)$ with entries in
  $\text{\tmtextrm{\tmtextup{D}}} [Y, Z]$ such that for every $(y, z) \in
  \text{\tmtextrm{\tmtextup{C}}}^{\ell + 1}$ such that $z \neq 0$, the matrix
  $M'_i (y, z)$ is the matrix of multiplication by~$zX_1, \ldots, zX_k$,
  expressed in the basis $\overline{\tmop{Mon}} ( \mathcal{G}) (z) .$
\end{itemize}{\hspace*{\fill}}{\medskip}

\begin{proof*}{Bit-size estimate:}
  Let $N = d_1 \cdots d_k, D = (d_1 + \cdots + d_k - k + 1)$ and let $\deg_Y
  (Q_i) \leq \lambda$ for $1 \leq i \leq k$. \ \ The entries of the matrix
  $M'_i$ of multiplication by $ZX_i$ in $\overline{\tmop{Mon}} ( \mathcal{G})
  (Z)$, have degrees in $Z$ bounded by $D$ and degrees in $Y$ bounded by $D
  \lambda$.
  
  When $\text{\tmtextrm{\tmtextup{D}}} = \mathbb{Z}$, the bitsize of \ the
  matrix $M'_i$ of multiplication by $ZX_i$ in $\overline{\tmop{Mon}} (
  \mathcal{G}) (Z)$ are bounded by
  \[ D (\tau + 2 \ell \tmop{bit} (D \lambda + 1) + \tmop{bit} (N)) - \ell
     \tmop{bit} (D \lambda + 1) - \tmop{bit} (N), \]
  where $\tau$ is a bound the bitsizes of the coefficients of the polynomials
  in $\mathcal{G}$.
\end{proof*}

\section{\label{sec:algebraic}Algebraic Case.}

In this section we prove Theorems \ref{12:the:algebraicsmallball} and
\ref{12:the:algebraicball}.

We first introduce some notation. Let $\text{\tmtextrm{\tmtextup{R}}}$ be a
real closed field. For any polynomial $P \in \text{\tmtextrm{\tmtextup{R}}}
[X_1, \ldots, X_k]$, let $\tmop{Zer}_b (P, \text{\tmtextrm{\tmtextup{R}}}^k)$
denote the union of the semi-algebraically connected components of $\tmop{Zer}
(P, \text{\tmtextrm{\tmtextup{R}}}^k)$ which are bounded over
$\text{\tmtextrm{\tmtextup{R}}}$. We denote by $\text{\tmtextrm{\tmtextup{R}}}
\langle \varepsilon \rangle$ the real closed field of algebraic Puiseux series
in $\varepsilon$ with coefficients in $\text{\tmtextrm{\tmtextup{R}}}$. The
elements of $\text{\tmtextrm{\tmtextup{R}}} \langle \varepsilon \rangle$ with
non-negative order constitute a valuation ring denoted
$\text{\tmtextrm{\tmtextup{R}}} \langle \varepsilon \rangle_b$. The elements
of $\text{\tmtextrm{\tmtextup{R}}} \langle \varepsilon \rangle_b$ are exactly
the elements of $\text{\tmtextrm{\tmtextup{R}}} \langle \varepsilon \rangle$
bounded over $\text{\tmtextrm{\tmtextup{R}}}$ (i.e. their absolute value is
less than a positive element of $\text{\tmtextrm{\tmtextup{R}}}$). We denote
by $\lim_{\varepsilon}$the ring homomorphism
from~$\text{\tmtextrm{\tmtextup{R}}} \langle \varepsilon \rangle_b$ to
$\text{\tmtextrm{\tmtextup{R}}}$ which maps $\sum_{i \in \mathbb{N}} a_i
\varepsilon^{i / q}$ to $a_0$. The mapping $\lim_{\varepsilon}$ simply
replaces~$\varepsilon$ by $0$ in a bounded Puiseux series.

\begin{proof*}{Proof of Theorem \ref{12:the:algebraicsmallball}:}
  In order to find a bound on the radius of a ball containing $\tmop{Zer}_b
  (Q, \text{\tmtextrm{\tmtextup{R}}}^k)$, it is enough to find an interval
  $[a, b]$ such that.
  \[ \tmop{Zer}_b (Q, \text{\tmtextrm{\tmtextup{R}}}^k) \subset [a, b] \times
     \text{\tmtextrm{\tmtextup{R}}}^{k - 1} . \]
  We are going to prove that it is possible to obtain such an interval from
  an interval $[a', b']$ such that the cylinder based on $[a', b']$ contains
  all the connected components bounded over $\text{\tmtextrm{\tmtextup{R}}}$
  of the zero sets of a convenient deformation of $Q$.
  
  Let $\zeta$ be a new variable. We define
  \begin{eqnarray*}
    Q_{\zeta}^+ & = & Q + \frac{\zeta}{d + 1} (X_1^{d + 1} + \cdots + X_k^{d +
    1}),\\
    Q_{\zeta}^- & = & Q - \frac{\zeta}{d + 1} (X_1^{d + 1} + \cdots + X_k^{d +
    1}) .
  \end{eqnarray*}

  Observe that $\tmop{Zer}_{} (Q_{\zeta}^+, \text{\tmtextrm{\tmtextup{R}}}
  \langle \zeta \rangle^k)$ (resp. $\tmop{Zer}_{} (Q_{\zeta}^-,
  \text{\tmtextrm{\tmtextup{R}}} \langle \zeta \rangle^k)$) is an hypersurface
  with isolated singular points since the ideal generated by
  \[ \frac{\partial Q_{\zeta}^+}{\partial X_1}, \cdots, \frac{\partial
     Q_{\zeta}^+}{\partial X_k}  \left( \tmop{resp} . \frac{\partial
     Q_{\zeta}^-}{\partial X_1}, \cdots, \frac{\partial Q_{\zeta}^-}{\partial
     X_k} \right) \]
  is zero-dimensional.
  
  Note that if $C$ is a bounded semi-algebraically connected component of
  $\tmop{Zer}_{} (Q, \text{\tmtextrm{\tmtextup{R}}}^k)$, there exists a finite
  number of semi-algebraically connected components $C_1, \ldots, C_c$ of
  $\tmop{Zer}_{} (Q_{\zeta}^+, \text{\tmtextrm{\tmtextup{R}}} \langle \zeta
  \rangle^k) \bigcup \tmop{Zer}_{} (Q_{\zeta}^-,
  \text{\tmtextrm{\tmtextup{R}}} \langle \zeta \rangle^k)$, bounded over
  $\text{\tmtextrm{\tmtextup{R}}}$ such that
  \[ C = \lim_{\zeta}  \left( C_1 \cup \ldots \cup C_c \right) . \]
  In order to see this first observe that
  \[ \tmop{Zer}_{} (Q_{\zeta}^+, \text{\tmtextrm{\tmtextup{R}}} \langle \zeta
     \rangle^k) \bigcup \tmop{Zer}_{} (Q_{\zeta}^-,
     \text{\tmtextrm{\tmtextup{R}}} \langle \zeta \rangle^k) = \tmop{Zer}_{}
     (Q_{\zeta}, \text{\tmtextrm{\tmtextup{R}}} \langle \zeta \rangle^k), \]
  where
  \[ Q_{\zeta} = Q^2 - \left(  \frac{\zeta}{d + 1} \right)^2 (X_1^{d + 1} +
     \cdots + X_k^{d + 1})^2, \]
  Moreover, the polynomial $(X_1^{d + 1} + \cdots + X_k^{d + 1})^2$ is
  non-negative everywhere in $\text{\tmtextrm{\tmtextup{R}}}^k$. Now apply
  Proposition 12.37 \ in {\cite{BPRbook2posted}}, after noting that by
  Proposition 12.35 $\lim_{\zeta}$ of a semi-algebraically connected component
  of {\color{red} }$\tmop{Zer}_{} (Q_{\zeta}, \text{\tmtextrm{\tmtextup{R}}}
  \langle \zeta \rangle^k)$ bounded over $\text{\tmtextrm{\tmtextup{R}}}$
  remains \ semi-algebraically connected and bounded over
  $\text{\tmtextrm{\tmtextup{R}}}$.
  
  This implies that, denoting by $\pi$ the projection to the $X_1$-axis,
  \begin{equation}
    \pi (C) = \lim_{\zeta} \left( \pi (C_1 \cup \ldots \cup C_c) \right) .
  \end{equation}

  Let $[a, b] = \pi (C)$, and $a_{\zeta}$ and $b_{\zeta}$ be the minimum and
  maximum of $\pi (C_1 \cup \ldots \cup C_c)$. It follows from (1) that
  $\lim_{\zeta} (a_{\zeta}) = a$, $\lim_{\zeta} (b_{\zeta}) = b$.
  
  In order to describe $a_{\zeta}$ and $b_{\zeta}$, we introduce
  zero-dimensional polynomial systems \ whose solutions correspond to critical
  points of $\tmop{Zer}_{} (Q_{\zeta}^+, \text{\tmtextrm{\tmtextup{R}}}
  \langle \zeta \rangle^k)$ (resp. $\tmop{Zer}_{} (Q_{\zeta}^-,
  \text{\tmtextrm{\tmtextup{R}}} \langle \zeta \rangle^k)$) in the
  $X_1$-direction, and compute the characteristic polynomials $\chi^+_{}
  (\zeta, T)$ (resp. $\chi^- (\zeta, T)$) of the multiplication by $\zeta X_1$
  (to avoid denominators). Since $a_{\zeta} \tmop{and} b_{\zeta}$ are extremal
  values of $\pi$ on $C_1 \cup \ldots \cup C_c$, they are roots of the
  polynomials $F^+ (\zeta, T) \in \text{\tmtextrm{\tmtextup{R}}} [\zeta, T]$
  and $F^- (\zeta, T) \in \text{\tmtextrm{\tmtextup{R}}} [\zeta, T]$ obtained
  by substituting $\frac{T}{\zeta} \tmop{to} T$ and multiplying by the minimum
  power of $\zeta$ necessary to avoid denominators. Finally, since $a_{\zeta}$
  and $b_{\zeta}$ are bounded over $\text{\tmtextrm{\tmtextup{R}}}$, $a$ and
  $b$ are roots of $f^+ (T) = F^+ (0, T)$ or $f^- (T) = F^- (0, T)$.
  
  We thus define the polynomial systems
  \begin{eqnarray*}
    \tmop{Cr} (Q_{\zeta}^+) & = & \left\{ (d + 1) Q_{\zeta^{}}^+ - \left( X_2 
    \frac{\partial Q_{\zeta}^+}{\partial X_2} + \cdots + X_k  \frac{\partial
    Q_{\zeta}^+}{\partial X_k} \right), \frac{\partial Q_{\zeta}^+}{\partial
    X_2}, \cdots, \frac{\partial Q_{\zeta}^+}{\partial X_k} \right\},\\
    \tmop{Cr} (Q_{\zeta}^-) & = & \left\{ (d + 1) Q_{\zeta^{}}^- - \left( X_2 
    \frac{\partial Q_{\zeta}^-}{\partial X_2} + \cdots + X_k  \frac{\partial
    Q_{\zeta}^-}{\partial X_k} \right), \frac{\partial Q_{\zeta}^-}{\partial
    X_2}, \cdots, \frac{\partial Q_{\zeta}^-}{\partial X_k} \right\} .
  \end{eqnarray*}
  Notice that the zero set of $\tmop{Cr} (Q_{\zeta}^+)$ (resp. $\tmop{Cr}
  (Q_{\zeta}^-)$) is the set of critical points on $\tmop{Zer}_{}
  (Q_{\zeta}^+, \text{\tmtextrm{\tmtextup{R}}} \langle \zeta \rangle^k)$
  (resp. $\tmop{Zer}_{} (Q_{\zeta}^-, \text{\tmtextrm{\tmtextup{R}}} \langle
  \zeta \rangle^k)$) in the $X_1$-direction, and that $\tmop{Cr}
  (Q_{\zeta}^+)$, $\tmop{Cr} (Q_{\zeta}^-)$ are both parametrized special
  Groebner basis with
  \[ d_1 = d + 1, d_2 = \cdots = d_k = d, \]
  and bitsizes of the coefficients bounded by
  \[ \tau + \tmop{bit} (d + 1) . \]
  According to the complexity analysis of Algorithm Parametrized Special
  Matrices of Multiplication, it follows that the matrix $M_{}^+$ (resp.
  $M^-$) of multiplication by $\zeta X_1$ has dimension
  \[ N = (d + 1) d^{k - 1}, \]
  and the bitsizes of its entries bounded by
  \[ \tau_1 = (k d + 1 - k + 1) (\tau + \tmop{bit} (d + 1) + \tmop{bit} (N))
     - \tmop{bit} (N) \]
  \[ = (k (d - 1) + 2) (\tau + \tmop{bit} (d + 1) + \tmop{bit} (N)) -
     \tmop{bit} (N) . \]
  So the characteristic polynomial $\chi^+_{} (\zeta, T)$ (resp. $\chi^-
  (\zeta, T)$) of $M^+$ \ (resp. $M^-$) is a polynomial in $\zeta$ and $T$
  with bitsize bounded by,
  \[ N (\tau_1 + \tmop{bit} (N) + k \tmop{bit} (d + 1) + 1) \leq N (kd + 2)
     (\tau + \tmop{bit} (d + 1) + \tmop{bit} (N)) \]
  and degrees in $Z$ bounded by $(k (d - 1) + 2) N$, using Proposition
  8.16{\color{red}  }of {\cite{BPRbook2posted}} and noting that the entries of
  $M$ are monomials in $Z$ of degrees bounded by $k (d - 1) + 2$.
  
  Thus, $a$ and $b$ are roots of a polynomial -either $f^+ (T)$ or \ $f^-
  (T)$- of degree at most $N$ and whose bitsizes are bounded by
  \[ N (kd + 2) (\tau + \tmop{bit} (N) + \tmop{bit} (d + 1)) . \]
  Using Cauchy bound (see {\cite{BPRbook2posted}}) we finally obtain that $a$
  and $b$ are bounded by
  \[ (N + 1) 2^{N (kd + 2) (\tau + \tmop{bit} (N) + \tmop{bit} (d + 1))} . \]
  The theorem follows immediately from this.
\end{proof*}


\begin{proof*}{Proof of Theorem \ref{12:the:algebraicball}:}
  For bounded connected components of $\tmop{Zer} (Q,
  \text{\tmtextrm{\tmtextup{R}}}^k$), we \ apply the previous theorem.
  
  To deal with unbounded connected components, let $\varepsilon$ be a new
  variable. We define
  
  \begin{equation}
    Q_{\varepsilon} = Q^2 + (\varepsilon (X_1^2 + \cdots + X_k^2) - 1)^2 .
  \end{equation}
  
  Notice that the extenstion to $\text{\tmtextrm{\tmtextup{R}}} \langle
  \varepsilon \rangle$ of every unbounded connected component of $\tmop{Zer}
  (Q, \text{\tmtextrm{\tmtextup{R}}}^k)$ meets $\tmop{Zer} (Q^2 + (\varepsilon
  (X_1^2 + \cdots + X_k^2) - 1)^2, \text{\tmtextrm{\tmtextup{R}}} \langle
  \varepsilon \rangle^k)$ and that $\tmop{Zer} (Q_{\varepsilon},
  \text{\tmtextrm{\tmtextup{R}}} \langle \varepsilon \rangle^k)$ is contained
  in the ball $\overline{B} (0, \varepsilon^{- 1 / 2})$. So $\overline{B} (0,
  \varepsilon^{- 1 / 2})$ \ intersects the extension to
  $\text{\tmtextrm{\tmtextup{R}}} \langle \varepsilon \rangle$ of every
  unbounded connected component of $\tmop{Zer} (Q,
  \text{\tmtextrm{\tmtextup{R}}}^k)$. We then replace $\varepsilon$ by a small
  enough positive $u \in \text{\tmtextrm{\tmtextup{R}}}$ and prove that
  $\overline{B} (0, u^{- 1 / 2})$ \ intersects every unbounded connected
  component of $\tmop{Zer} (Q, \text{\tmtextrm{\tmtextup{R}}}^k)$.
  
  Noting that $Q_{\varepsilon}$ is everywhere non-negative, we can proceed as
  in the proof of Theorem 1 and take
  \begin{eqnarray*}
    Q_{\varepsilon, \zeta} & = & Q_{\varepsilon} - \frac{\zeta}{d'} (X_1^{d'}
    + \cdots + X_k^{d'} + d' (X_{1^{}}^2 + \cdots + X_{k^{}}^2) + k (d' +
    1)),\\
    &  & \\
    \tmop{Cr} (Q_{\varepsilon, \zeta}) & = & \left\{ d' Q_{\varepsilon, \zeta}
    - \left( X_2  \frac{\partial Q_{\varepsilon, \zeta}}{\partial X_2} +
    \cdots + X_k  \frac{\partial Q_{\varepsilon, \zeta}}{\partial X_k}
    \right), \frac{\partial Q_{\varepsilon, \zeta}}{\partial X_2}, \cdots,
    \frac{\partial Q_{\varepsilon, \zeta}}{\partial X_k} \right\},
  \end{eqnarray*}
  with $d' = \sup (2 (d + 1), 6)$.
  
  {\color{red} }Note that for every unbounded connected component $D$ of
  $\tmop{Zer} (Q, \text{\tmtextrm{\tmtextup{R}}}^k)$, the elements of
  $\lim_{\zeta} (\tmop{Zer} ( \tmop{Cr} (Q_{\varepsilon, \zeta}^{}),
  \text{\tmtextrm{\tmtextup{R}}} \langle \varepsilon \rangle \langle \zeta
  \rangle))$ meet $\tmop{Ext} (D, \text{\tmtextrm{\tmtextup{R}}} \langle
  \varepsilon \rangle)$ by Proposition 12.37 of {\cite{BPRbook2posted}}.
  
  Moreover, $\tmop{Cr} (Q_{\varepsilon, \zeta} $) is a paramaetrized special
  Groebner basis with
  \[ Z = \zeta, Y_1 = \zeta, Y_2 = \varepsilon, d_1 = d', d_2 = \cdots = d_k =
     d' - 1, \ell = 2, \lambda = 2, \]
  and the bitsizes of the coefficients bounded by
  \[ 2 \tau + k \tmop{bit} (d + 1) + \tmop{bit} (d') + 1. \]
  According to the complexity analysis of Algorithm Parametrized Special
  Matrices of Multiplication, it follows that the matrix $M_i$ of
  multiplication by $\zeta X_i$ has dimension
  \[ N = d' (d' - 1)^{k - 1}, \]
  and the bitsizes of its entries is bounded by
  \[ D (2 \tau + k \tmop{bit} (d + 1) + \tmop{bit} (d') + 1 + 4 \tmop{bit}
        (2 D + 1) + \tmop{bit} (N)) - 2 \tmop{bit} (2 D + 1) - \tmop{bit} (N),
  \] 
  where
  \begin{eqnarray*}
    D & = & d' + (k - 1) (d' - 1) - k + 1 = k d' - 2 (k - 1),
  \end{eqnarray*}
  while the degree in $\varepsilon, \zeta$ of its entries is bounded by $2 D$.
  
  For every $j$, denote by $L_j$ the matrix of mutiplication by the linear
  form $\zeta (X_1 + j X_2 + \cdots + j^{k - 1} X_k)$, by $\chi (j,
  \varepsilon, \zeta, T)$ its characteristic polynomial and by $G (j,
  \varepsilon, \zeta, T)$ the polynomial obtained by substituting
  $\frac{T}{\zeta} \tmop{to} T$ in $\chi (j, \varepsilon, \zeta, T)$  and
  multiplying by the minimum power of $\zeta$ necessary to avoid denominators,
  and by $g (j, \varepsilon, T)$ the polynomial $G (j, \varepsilon, 0, T)$. It
  follows from {\cite{BPRbook2posted}} that there exists $0 \leq j \leq (k -
  1) N^2$, such that every point $x (\varepsilon)$ of $\lim_{\zeta}
  (\tmop{Zer} ( \tmop{Cr} (Q_{\varepsilon, \zeta}^{}),
  \text{\tmtextrm{\tmtextup{R}}} \langle \varepsilon \rangle \langle \zeta
  \rangle))$is of the form $r (\varepsilon, t (\varepsilon))$ where $t
  (\varepsilon)$ is a root of $g (j, \varepsilon, T)$ and $r (\varepsilon, T)$
  is a rational function with denominator a derivative of $g (j, \varepsilon,
  T)$. Finally, for every unbounded connected component $D$ of $\tmop{Zer} (Q,
  \text{\tmtextrm{\tmtextup{R}}}^k)$, there is a root $t (\varepsilon)$ of $g
  (j, \varepsilon, T)$ and a rational function $r (\varepsilon, T)$ with
  denominator a derivative of $g (j, \varepsilon, T)$ such that $r
  (\varepsilon, t (\varepsilon)) \in \tmop{Ext} (D,
  \text{\tmtextrm{\tmtextup{R}}} \langle \varepsilon \rangle^{})$.
  
  The matrix $M$ of mutiplication by the linear form $\zeta (X_1 + j X_2 +
  \cdots + j^{k - 1} X_k)$ \ has entries with bitsizes bounded by $2 D \tau +
  \rho + \sigma$, with
  \begin{eqnarray*}
    \rho & = & D (k \tmop{bit} (d + 1) + \tmop{bit} (d') + 1 + 4 \tmop{bit} (2
    D + 1) + \tmop{bit} (N)) - 2 \tmop{bit} (2 D + 1),\\
    \sigma & = & (2 k - 3) \tmop{bit} (N) + k \tmop{bit} (k) .
  \end{eqnarray*}
  So the characteristic polynomial, $\chi (j, \varepsilon, \zeta, T)$ of $M$
  is a polynomial in $\varepsilon, \zeta, T$ with degree in $T$ bounded by
  $N$, degree in $\varepsilon, \zeta$ bounded by $2 D N$, and bitsize bounded
  by
  \[ \tau' = 2 ND \tau + N (\rho + \rho') \]
  with
  \begin{eqnarray*}
    \rho' & = & (2 k - 2) \tmop{bit} (N) + k \tmop{bit} (k) + 2 \tmop{bit} (2
    DN + 1) + 1,
  \end{eqnarray*}
  using Proposition 8.16 of {\cite{BPRbook2posted}}.The same estimate holds
  for the bitsize of $g (j, \varepsilon, T)$.
  
  Now let $u_0 \in \text{\tmtextrm{\tmtextup{R}}}$, with $u_0 > 0$, be such
  that the number and multiplicities of the real roots of $g (j, u, T)$ stay
  constant for all $u \in (0, u_0)$ and denote by $t (u)$ the root of $g (u,
  T)$ having the same number as $t (\varepsilon)$ as a root of $g (j_{},
  \varepsilon, T)$. Then for every point $x (\varepsilon)$ of $\lim_{\zeta}
  (\tmop{Zer} ( \tmop{Cr} (Q_{\varepsilon, \zeta}^{}),
  \text{\tmtextrm{\tmtextup{R}}} \langle \varepsilon \rangle \langle \zeta
  \rangle))$, such that $x (\varepsilon) = r (\varepsilon, t (\varepsilon))$,
  the function $r (u, t (u))$ is defined from $(0, u_0$) to $\tmop{Zer} (Q,
  \text{\tmtextrm{\tmtextup{R}}}^k)$. The graph of this function is connected
  and intersect \ $D$, since $r (\varepsilon, t (\varepsilon)) \in \tmop{Ext}
  (D, \text{\tmtextrm{\tmtextup{R}}} \langle \varepsilon \rangle^{})$.
  
  Let $\mathcal{A (\varepsilon)}$ be the set of all subresultants of $g
  (\varepsilon, T)$ and $g^{(\ell)} (\varepsilon, T)$, $1 \leq \ell \leq N -
  1$, with respect to the variable $T$. From the definition of the
  subresultants (see {\cite{BPRbook2posted}}), the polynomials in $\mathcal{A
  (\varepsilon)}$ have degrees in $\varepsilon$ bounded by
  \[ 2 D N (2 N - 1) \]
  and bitsizes bounded by
  \[ (2 N - 1) \tau' + N^2 \tmop{bit} (N + 1) . \]
  Choosing $u_0$ smaller that the smallest positive root of the polynomials in
  $\mathcal{A (\varepsilon)}$, the number and multiplicities of the real roots
  of $g (j, u, T)$ stay constant for all $u \in (0, u_0)$ by using the
  properties of subresultants.
  
  Finally, applying Cauchy bound (see {\cite{BPRbook2posted}}). we see that we
  can choose the rational number $u_0$ of bitsize bounded by
  \[ (2 D N (2 N - 1) + 1) 2^{\left. (2 N - 1) \tau' + N^2 \tmop{bit} (N + 1
     \right)} . \]
\end{proof*}

\section{\label{sec:semi-algebraic}Semi-algebraic case}

\begin{proof*}{Proof of Theorem \ref{13:the:diam}:}
  We first observe that given a bounded semi-algebraically connected
  component $D$ of a basic semi-algebraic set defined by $P_1 \geq 0, \ldots,
  P_s \geq 0$, and let $w \in \text{\tmtextrm{\tmtextup{R}}}$ be an extremal
  value (either maximum or minimum) of the $X_1$-co-ordinate realized on $D$.
  Then, there exists $\{i_1, \ldots, i_m \} \subset \{1, \ldots, s\}$ and a
  bounded semi-algebraically connected component $C$ of the algebraic set
  $\tmop{Zer} (\{P_{i_1}, \ldots, P_{i_s} \},
  \text{\tmtextrm{\tmtextup{R}}}^k)$ such that $C \subset D$ and $w$ is the
  extremal value of the $X_1$-co-ordinate realized on $C$. Indeed, let $W
  \subset D$ be the set of points of $D$ with their first co-ordinate equal to
  $w$. For any point $x \in W$ let $\mathcal{P}_x =\{P \in \mathcal{P} \mid P
  (x) = 0\}$. We choose $x \in W$ such that $\mathcal{P}_x =\{P_{i_1}, \ldots,
  P_{i_m} \}$ is maximal with respect to inclusion. Let $C$ be the connected
  component of $\tmop{Zer} (\{P_{i_1}, \ldots, P_{i_s} \},
  \text{\tmtextrm{\tmtextup{R}}}^k)$ which contains $x$. Then, $C \subset D$
  by the maximality of $\mathcal{P}_x =\{P_{i_1}, \ldots, P_{i_m} \}$ and
  clearly $C$ is bounded since $D$ is bounded, and moreover $w$ is an extremal
  value of $C$.
  
  We now apply Theorem \ref{12:the:algebraicsmallball} above, noting that for
  any subset $\mathcal{P}' \subset \mathcal{P}$, the bitsizes of the
  coefficients of the polynomial $\sum_{P \in \mathcal{P}'} P^2$ is bounded by
  $2 \tau + k \tmop{bit} (d + 1) + \tmop{bit} (s)$, and its degree is bounded
  by $2 d$. Applying Theorem \ref{12:the:algebraicsmallball} we obtain that
  \[ w \leq (N + 1) 2^{N (2 kd + 2) (2 \tau + \tmop{bit} (N) + (k + 1)
     \tmop{bit} (d + 1) + \tmop{bit} (s))}, \]
  where
  \[ N = (2 d + 1) (2 d)^{k - 1} . \]
\end{proof*}

\begin{proof*}{Proof of Theorem \ref{13:the:saball}:} 
  Since every semi-algebraically connected
  component of the realization of a weak sign condition on $\mathcal{P}$ must
  contain a connected component of some algebraic set $\tmop{Zer}
  (\mathcal{P}', \text{\tmtextrm{\tmtextup{R}}}^k)$, where $\mathcal{P}'
  \subset \mathcal{P}$, it suffices to apply Theorem
  \ref{12:the:algebraicball} to obtain an upper bound on the radius of a ball
  guaranteed to meet all such components. Note that for any subset
  $\mathcal{P}' \subset \mathcal{P}$, the bitsizes of the coefficients of the
  polynomial $\sum_{P \in \mathcal{P}'} P^2$ is bounded by $2 \tau + k
  \tmop{bit} (d + 1) + \tmop{bit} (s)$, and its degree is bounded by $2 d$.
  Note also that we can use directly $\sum_{P \in \mathcal{P}'} P^2$ without
  squaring in (2). The theorem is then a straightforward consequence of
  Theorem \ref{12:the:algebraicball}.
\end{proof*}

\end{document}